\documentclass[american,runningheads,envcountsect]{llncs}
\usepackage[utf8]{inputenc}
\usepackage[T1]{fontenc}
\usepackage{babel}
\usepackage{url}
\usepackage{tikz}
\usepackage{standalone} 
\usepackage[kerning=true]{microtype}
\usepackage{amsmath,amssymb,amsfonts}
\usepackage{subcaption}

\usepackage{tabularx}
\usepackage[backend=bibtex,style=numeric-comp,doi=false,isbn=false,%
giveninits=true,url=false]{biblatex}
\usepackage{mathtools}
\usepackage{graphicx}
\usepackage{paralist}
\usepackage{float}

\usepackage{textcomp}
\usepackage[colorlinks,citecolor=blue,urlcolor=black,hidelinks,linktocpage]{hyperref}
\usepackage{cleveref}

\usepackage{xcolor}
\usepackage{blindtext}

\usepackage{todonotes}
\presetkeys{todonotes}{color=blue!5}{}

\usepackage{booktabs}
\usepackage{multirow}

\let\cref\Cref
\begin{document}
\title{Topological Indoor Mapping through WiFi Signals}


\institute{%
  Witheld Institute
  \email{\{withheld\}@witheld}
}

\author{Bastian Schaefermeier\inst{1,2}\orcidID{0000-0002-5556-7608}%
  \and Gerd~Stumme\inst{1,2}\orcidID{0000-0002-0570-7908}%
  \and Tom~Hanika\inst{1,2}\orcidID{0000-0002-4918-6374}%
}
\authorrunning{Schaefermeier et al.}

\institute{%
  Knowledge \& Data Engineering Group,
  University of Kassel, Germany
  \and
  Interdisciplinary Research Center for Information System Design
  (ITeG), U Kassel\\[0.5ex]
  \email{\{schaefermeier, stumme, tom.hanika\}@cs.uni-kassel.de}
  \vspace{-0.8cm}
}

\maketitle
\begin{abstract}
The ubiquitous presence of WiFi access points and mobile devices
 capable of measuring WiFi signal strengths allow for real-world applications 
 in indoor localization and mapping. In particular, no additional infrastructure is required.
Previous approaches in this field were, however, often hindered by problems 
such as effortful map-building processes, changing environments and hardware differences.
We tackle these problems focussing on \emph{topological maps}. These represent
discrete locations, such as rooms, and their relations, e.g., distances and transition frequencies.
In our unsupervised method, we employ WiFi signal strength distributions, dimension reduction and clustering. 
It can be used in settings where users carry mobile devices and follow their normal routine.
We aim for applications in short-lived indoor events such as conferences.
\end{abstract}

\section{Introduction}

Navigating unfamiliar indoor venues, such as conference buildings or
fair grounds, is a reoccurring task for many individuals. A map that
depicts the particular venue layout is a welcome support which,
however, has to be prepared beforehand. Moreover, especially at
events, such as scientific conferences, this map can be dynamic and
require continuous modifications. Hence, techniques to automatically
derive maps for indoor venues are useful for a wide range of events.

Assuming that almost every person attending an event has a WiFi
enabled smartphone and that WiFi access points (APs) are ubiquitous,
in almost all indoor venues, it is natural to develop mapping
procedures that take advantage of these facts. WiFi Signals have often
been used for indoor localization and mapping in recent
years~\cite{wifi_survey, indoorloc_survey}. Since conventional smartphones can
measure signal strengths, in theory, no additional infrastructure or
mapping devices are
required. 
However, there are two main problems that arise in such a scenarios:
first, it is hard to achieve good localization accuracy through WiFi
signals. The reason for this are disturbances of WiFi signals, e.g.,
changes in the environment or humans blocking signals. Additionally,
received signal strength indicator (RSSI) measurements are device
dependent due to different hardware
specifications. 
Secondly, it requires much effort to map all relevant sites, i.e.,
physical locations and to represent them via WiFi signal strengths
(called fingerprints). Usually such fingerprints need to be collected
manually before the so-called \emph{online phase}. Approaches for
automatic site survey exist, yet, they suffer from other
difficulties. They often require specific user trajectories and
behavior, e.g., to hold a smartphone in a particular way. More
sophisticated methods also require modifications of the WiFi
infrastructure by modifying AP software.


In our present work we propose a (lightweight) method for
\emph{topological indoor mapping}~\cite{topomap} from RSSI data
recorded by smartphones. We figure that accurate topological maps are
favored over inaccurate metric maps, in particular within the scope of
short-lived indoor events. Hence, our method derives graphs from RSSI
data, in which nodes represent (important) locations, e.g., seminar
rooms or a refreshments table, and edges represent transitions between
these locations (e.g., hallways or stairways). We argue that modeling
locations in that manner is similar to the way that humans perceive
and refer to their immediate
environment~\cite{Shi_2010}. 
Apart from the conceptual benefits, our hence coarse-grained modeling
comes with the practical advantage that accurate localization is
easier to achieve. 
We cope with the hardware dependence of our measurements as well as
signal disturbances by employing RSSI distributions of locations (RSSI
likelihoods), which are less error prone than single measurements. Our
contributions can be summarized as follows:
\begin{inparaenum}[a)]
\item We present a lightweight method for building topological indoor
  maps from WiFi data. 
\item We experiment on different clustering- and layout techniques,
  from which we can derive recommendations for similar applications.
\item We demonstrate that neither special user behavior nor technical
  changes of the WiFi infrastructure are required to achieve
  practically usable results, as demonstrated by two experiments
  conducted in real-world scenarios.  
\end{inparaenum}



\section{Related Work}
An overview on WiFi localization can be gained in the survey articles by He and Chan~\cite{wifi_survey} and Zafari et al.~\cite{indoorloc_survey}. Notable early approaches with a lasting influence on the field are Horus \cite{horus2005} and Radar \cite{radar}. In both methods, samples of WiFi received signal strength indicators (RSSI) for different locations are collected beforehand in a so-called \emph{offline phase}. The task in the \emph{online phase} is to perform localization based on a new set of RSSI measurements. Radar achieves this through $k$-nearest-neighbors in a vector space of RSSI signals. Horus represents locations as probability distributions and solves localization using a maximum-likelihood approach. Both methods require manual effort in building WiFi maps in advance by recording labeled RSSI samples (called \emph{fingerprints}).

To circumvent manual mapping, \emph{simultaneous localization and mapping} (SLAM) approaches were adopted to WiFi mapping \cite{signalslam, conf/ijcai/FerrisFL07, conf/aaai/XiongT17}. These methods are based on pedestrian dead reckoning, which requires additional sensor data, e.g., from accelerometer and gyroscope. However, they were only successful in small controlled scenarios and found to be very sensitive to the accuracy and precision of the used smartphone sensors, which may vary in a practical scenario. 

WiFi-based topological indoor maps were created by Shin and Cha~\cite{topomap}. Places are represented as nodes, paths as edges in a graph. In the resulting visualizations, nodes were required to be laid out manually. Hence, much of the topology is not captured in the automatic process. Places are points in a coordinate grid. In contrast to this, our aim is that each node represents a meaningful abstract location, such as a room.
The problem of creating topological indoor maps in our sense was introduced by Schaefermeier et al.~\cite{wifidistances}. In that work, research
was focussed on the first steps, namely location representations and distance measures. In this work, we build upon this to create complete topological maps.

\section{Problem Statement}
\paragraph{Task and Requirements}
Our overall goal is to automatically infer topological maps of
physical venues from signal strength measurements of WiFi access
points (APs). In detail, in our setup people shall carry WiFi enabled
smartphones while conducting their business within a physical space,
e.g., attending talks in a conference building. These smartphones,
equipped with a \emph{topological navigation application}, passively
measure the different signal distributions at different places.  A
therefrom computed \emph{topological map} should reflect
\begin{inparaenum}[a)]
\item which physical locations do exist, and
\item how these locations are related to each other. 
\end{inparaenum}
Physical locations can be, e.g., a kitchen in an office or a seminar
room in a conference building.

We require any approach to achieve our goal to be:
\begin{inparadesc}
\item[Independent,]i.e., our method should work without any additional
  infrastructure at the physical locations. 
\item[Automatic,]i.e., our method should not require a specific user
  behaviour (e.g., pointing the smartphone in walking direction).
\item[Effortless,]i.e., no user interaction or manual place
  annotation should be needed.
\item[Lightweight,]i.e., we aim at using as few data as needed and, moreover, we intend to
keep the model complexity as low as possible.
\end{inparadesc}
To discriminate sets of WiFi measurements, one needs, first, a
representation of such measurements, called \emph{location fingerprint},
and secondly, a dissimilarity function or \emph{distance} between any
two representations.
%


We employ \emph{received signal strength indicator} (RSSI)
measurements that are often represented as vectors, where each
component represents the RSSI of a specific AP measured at particular
point in time~\cite{radar}. In this work we grasp them as probability
distributions of RSSI values that are conditioned on locations. This
approach is less prone to random fluctuations of single
measurements. Moreover, it allows naturally for inferring localization
of previously unseen measurements by maximum likelihood and
determining a confidence that a measurement was made at a particular
location. Hence the name \emph{RSSI likelihoods}.
%


\paragraph{Formal Problem Definition}
\label{sec:formalProb}

The data basis is constituted by timestamped WiFi observations, e.g., recorded by a
smartphone. We draw on the relation between signal strength and
physical distance to an AP. Each AP is uniquely identified through a
basic service set identifier (BSSID).
\begin{definition}[WiFi data set]
  We call \( W \subseteq \mathbb{N} \times D \times B \times R \)
  \emph{WiFi data set}, where $\mathbb{N}$ represents timestamps, $D$
  is a set of devices, $B$ a set of BSSIDs and $R=[-100, -10]\cap
  \mathbb{Z}$ a range of RSSI values. A WiFi observation $o\coloneqq
  (t, d, b, r ) \in W$ contains the RSSI value $r$ of WiFi access
  point $b$ sensed at timestamp $t$ by device $d$. For all
  $(t,d,b,s_1), (t,d,b,s_2) \in W$ it shall hold that $s_1 =
  s_2$.
\end{definition}


Let the equivalence relation $\check{L}\subseteq W \times W$ reflect
the \textsl{true} physical locations, i.e., any two elements of
$\check{L}$ represent two distinct physical locations. 
The classes $[o]_{\check L}= \{o' \in W \mid (o, o') \in \check L \}$
give rise to the partition $\check{\mathcal{P}}=W/\check L$. 
The \emph{computational problem} in our work is to derive an approximation
$\mathcal{P}$ of $\check{\mathcal{P}}$, that also allows to lay out
useful topological maps. 
Note, $L \coloneqq \{ (u,v) \mid \exists P \in \mathcal{P}: u \in P \land v\in P \}$.
For the approximation we investigate different layout
methods and clustering techniques. Furthermore, we enrich our data
basis for the map building purpose with an \emph{acceleration data set}, i.e., a
relation $A \subset \mathbb{N} \times D \times \mathbb{R}$. An element
$(t,d,a) \in A$ represents an \emph{acceleration observation}, where
$a\coloneqq\|\mathbf{a}\|$ for an acceleration vector $\mathbf{a}\in
\mathbb{R}^3$. This allows us to identify time intervals in which a
device is stationary, an information from which we can infer a
``preclustering'' of $W$.

\section{Method}

\usetikzlibrary{shapes,arrows}
\tikzstyle{proc} = [rectangle, draw, minimum height=2.5em, text width=6.3em, text centered]
\tikzstyle{res} = [minimum height=2em, text width=5.5em, text centered]
\tikzstyle{line} = [draw, -latex']

\begin{figure}[b]
\begin{center}
\scalebox{0.7}{
\begin{tikzpicture}[node distance = 6.5em, auto, scale=0.10]
\node [proc] (mode) {Motion Mode Segmentation};
\node [res, left of=mode, node distance=3.5cm] (wifi) {WiFi Data $W$};
\node [res, below of=wifi, node distance=1cm] (acc) {Acceleration Data $A$};
\node [proc, below of=mode, node distance=3cm] (like) {Likelihood Estimation};
\node [proc, right of=like, node distance=4.5cm] (dist) {Distance Calculation};
\node [proc, node distance=4cm, right of=dist] (layout) {Layout};
\node [proc, above of=layout, node distance=3cm] (clustering) {Clustering};
\node [res, right of=clustering, node distance=3.5cm] (result) {Segment Clustering (Locations)};
\path [line] (acc) -- (mode);
\path [line] (wifi) -- (mode);
\path [line] (mode) -- node[res]{Stationary WiFi Segments}  (like);
\path [line] (like) -- node[res]{RSSI Likelihoods} (dist);
\path [line] (dist) -- node[res]{Distance Matrix} (layout);
\path [line] (layout) -- node[res]{Segment Coordinates}  (clustering);
\path [line] (clustering) -- (result);
\end{tikzpicture}
}
\end{center}
\caption{Topological indoor mapping procedure. Each step of our method is depicted as a block. Labels outside blocks refer to their inputs and outputs.}
\label{fig:method}
\end{figure}
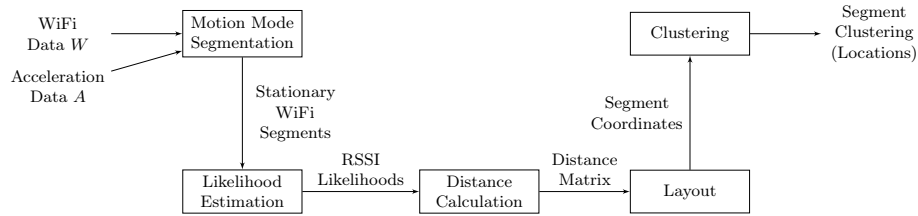
Our method, as depicted in \cref{fig:method}, is a sequence of steps
that executed in their order lead to the final set of locations. These
steps can be seen as conceptual building blocks, which can be realized
in different manners. The steps \emph{motion mode segmentation},
\emph{likelihood estimation} and \emph{distance calculation} have been
introduced and evaluated in previous work \cite{wifidistances}. For
completeness, we briefly sum up the recommended methods. For the step
\emph{layout}, we will do an empirical comparison of different methods
in \cref{sec:experiments}. For the step \emph{clustering} we select a
method based on theoretical and practical considerations.

\subsection{Motion Mode Segmentation}\label{sec:motionmode}
The first step of our method exclusively makes use of the acceleration data.
In detail, we classify the motion mode of a person into two classes, \emph{stationarity} or \emph{movement}.
We achieve this by applying a threshold to the energy \cite{motion_survey} calculated over short, consecutive time windows. 
We then summarize all consecutive windows with the same motion mode in a single interval $[t_i, t_{i+1})$, called \emph{motion mode segment}. We call $\sigma_d = (t_0, t_1, \ldots, t_n)$ the motion mode segmentation of $d$.
The idea behind this segmentation process is to combine WiFi observations recorded in segments of stationarity. For one such segment, we exploit the fact that all WiFi observations must have been recorded at a single location. Our reasoning is that single WiFi observations may often be distorted by various signal disturbances, while combining several can mitigate such problems. As a further benefit, we save considerable amounts of battery power by implementing the segmentation directly on the smart device. The reason is that after some time in stationarity we can stop recording further WiFi data, since the signal distribution can already be estimated from the samples recorded up to that point. We set the stopping threshold to five minutes, since this was sufficient for the estimation process.


\subsection{Likelihood Estimation}
\label{likeli}
For each stationary motion segment of a device $d$ in an interval
$[t_i, t_{i+1})$, we call the corresponding WiFi observations
$W_d(t_i) \subseteq W$ a \emph{stationary WiFi segment}. For each
stationary WiFi segment, we represent the RSSI observations made from
each access point $b_i \in B$ as a conditional probability
distribution $p_i(r_k \mid L_{d,t_j})$. This distribution models the
likelihood of receiving RSSI values $r_k$ when being at location
$L_{d,t_j}$. Therefore we call one such distribution \emph{RSSI
  likelihood} as an abbreviation.  To estimate RSSI likelihoods, we
use kernel density estimation (KDE) with a Gaussian kernel and a
constant bandwidth derived from background knowledge. We found
confirmation that this method improves WiFi localization
\cite{implicationsDiversity}. Park et al.~\cite{implicationsDiversity}
also found that KDE mitigates the \emph{device diversity problem},
i.e., localization problems caused by differences in hardware, which
in turn lead to different received signal strengths.  Different to the
aforementioned work, we also model the probability of AP invisibility
in our distributions. Each time signal strength is measured by a
device $d$ at time $t$ and nothing received from AP $b_i$, we add a
pseudo-observation $(t,d,b_i,-100)$ to the WiFi data.


\subsection{Distance Calculation} \label{sec:distcalc}
Let $P$ and $Q$ be the cumulative distribution functions of two univariate probability distributions $p$ and $q$ (representing RSSI likelihoods). We calculate their Earth Mover's Distance (EMD) as $EMD(P,Q)=\int{|P(x)-Q(x)|\mathrm dx}$.
EMD has several beneficial properties for calculating dissimilarities between RSSI likelihoods. EMD is a true metric and therefore coincides with our view of distances between physical locations in the real world, i.e., in $\mathbb{R}^2$ or $\mathbb{R}^3$. Furthermore it provides a smooth distance estimate and hence robustness against random differences in the sampled RSSI observations. It has been shown in previous work \cite{wifidistances}, that between distributions separated by a large gap, EMD approximates the differences between their means. For overlapping distributions, on the other hand, it captures more subtle differences, e.g., between their variances and kurtoses.
We combine the distances over all access points using the $L_1$ norm. Since EMD returns strictly positive values, the distance between the multivariate RSSI likelihoods $p$ and $q$ becomes the sum $d(p,q) = \sum_{i=1}^{k}{EMD(p_i, q_i)}$ over the univariate RSSI likelihoods $p_i$ and $q_i$ of each AP $b_i \in B$. We calculate distances between each pair of RSSI likelihoods and hence obtain a quadratic distance matrix $(d_{ij})$.

\subsection{Layout}
Through the calculated distances between WiFi segments, we gain information about their \emph{relative} positions in some abstract signal space. Clustering algorithms, which we use in further steps, in the general case, depend on \emph{absolute} positions, i.e., coordinates. Therefore we lay out segments in a $p$-dimensional space using three different methods (also called \emph{embedding algorithms}): Multidimensional Scaling (MDS), t-distributed Stochastic Neighborhood Embedding (t-SNE) and Uniform Manifold Approximation and Projection (UMAP). We chose MDS and t-SNE for being arguably the most widely used methods and UMAP as being a more recent state-of-the-art reported to perform well in, e.g., clustering \cite{umapclustering}.

Our intuition with embeddings is that the WiFi data is situated in a
two- (for a single floor) or three-dimensional (for several floors)
subspace of the original signal space. In this subspace, topological
relations between physical locations are retained, i.e., measurements
that are close in signal space arise from physically close locations. As a side-effect, we conjecture that because input distances can only approximately be retained, some noise and distortions are removed. 
We may note that the three methods are sometimes regarded as methods for visualization, and there is some dispute on the appropriateness of applying them before clustering. However, they have successfully been used in such scenarios before \cite{tsneclustering, umapclustering} and there have been some theoretical arguments for this application \cite{linderman2017tsne, linderman2019tsne}. 
Finally, we will use them for visual comparisons of the discovered
layouts to the ground truth. Having different properties, the three
methods may reveal and emphasize different aspects of the topological
relations between locations.

\paragraph{Multidimensional Scaling (MDS)}
Multidimensional scaling \cite{mds} finds coordinates in $\mathbb{R}^p$, such that the input distances $d_{ij}$ between the input pairs are retained as closely as possible. More formally, it finds coordinates $x_i \in \mathbb{R}^p$ for a chosen $p \in \mathbb{N}$ which minimize the so-called \emph{stress}, i.e., the function $S(x_1, x_2, \cdots, x_n) = \sum_{i \neq j}{ (d_{ij}-{\Vert x_i-x_j\Vert })^2}$.
In the case where the $d_{ij}$ arise from a metric, such as the EMD,
the distance matrix $(d_{ij})$ is symmetric and a triangle matrix is a
sufficient representation. The stress can thus be written as $\sum_{i
  < j}{ (d_{ij}-{\Vert x_i-x_j\Vert })^2}$ and can be minimized using the SMACOF algorithm \cite{smacof}.

\paragraph{t-Distributed Stochastic Neighborhood Embedding (t-SNE)}
t-SNE \cite{tsne_jmlr, tsne_aistats} finds layouts such that the local neighborhoods of points are preserved and accentuated. This makes resulting data points easier to cluster, because the distance to points from other clusters is emphasized. 
However, depending on the chosen hyper-parameters, t-SNE may also
overemphasize random differences in distances and hence lead to too many clusters. t-SNE tries to preserve affinities, which are joint-probabilities based on Gaussian distributions in the input space and based on Cauchy distributions in the output space.

\paragraph{Uniform Manifold Approximation and Projection (UMAP)}
This method~\cite{umap} creates a high- and a low-dimensional fuzzy topological representation of the input data set. The low-dimensional representation is found by minimizing a cross-entropy measure between the input and output fuzzy sets through stochastic gradient descent. 
An advantage of UMAP is that it produces stable results for different samples from the same data set while providing a low runtime. Finally, the resulting point sets are often densely concentrated facilitating clustering. 

\subsection{Clustering} \label{sec:clustering} We apply clustering
algorithms to the coordinates derived from one of the layout
methods. By a \emph{clustering} $\mathcal{C}$ of a set $X := \{x_1,
x_2, \ldots, x_n\}$ we denote a partition of $X$, i.e., $\mathcal{C}$
is a set $\mathcal{C}\coloneqq\{C_1, C_2, \ldots C_k\}$ of $k$
non-empty \emph{clusters} $C_i$ s.t. for all $i \neq j$ it holds that
$C_i \cap C_j = \emptyset$ and ${\bigcup}_{i=1}^k{C_i} = X$. In our
case, data points represent stationary WiFi segments of a device. The
partition elements (i.e., sets of WiFi segments) represent locations
found in our WiFi data set $W$. Thus, we solve two tasks through
clustering: \emph{location identification} and \emph{device
  localization}.

Our expectations for a clustering algorithm are:
\begin{inparaenum}[a)]
\item Stability, i.e., the same data set should lead to the
same clustering when repeatedly executing the algorithm. Small
perturbations in the data set should return a similar clustering.
\item Agreement, i.e., the found clusters should match our
  expectations of physical locations in the real world.
\end{inparaenum}
We compared these properties over various clustering algorithms and
experimented on their behavior. Many algorithms, however, suffer from
strong variances in the results. This is for two reasons: first, they
are sensitive to outliers. Second, they have elements of randomness,
which strongly affect the outcome and may lead to different results in
repeated runs. As a further problem, they often require the user to
set the number of clusters in advance, which, however, for our problem
is unknown. In our case, this translates to finding the right number
of locations (in agreement with the number of physical locations).

Based on these considerations, we selected the HDBSCAN
algorithm~\cite{ref_hdbscan}, an extension of DBSCAN. This algorithm
estimates the density of the underlying point distribution based on
the distance of each sample in the data set to its $k$-th nearest
neighbor. HDBSCAN determines the number of cluster based on the
density of the data and provides robustness detecting outliers.

Since device localization is done a-posterior, i.e., after the whole
data set has been collected, the clustering is comparable to what is
often called the \emph{offline phase} in localization methods. In this
phase, location samples are collected that are later used for
reference. Hence, based on the clustered WiFi data, we can easily
perform \emph{online} (i.e., real-time) device localization in
topological maps through a classification method such as, e.g., a
maximum likelihood approach.

\subsection{Graph Representation}\label{sec:graphrepresentation}
Once we have determined a partition $\mathcal{P}$ of our WiFi data set $W$ through clustering, we can order the locations of devices in time to receive their trajectories.
\begin{definition}[Trajectory of a Device]
A trajectory $T_d$ of a device $d$ in $W$ with partition $\mathcal{P}$ is a sequence $T_d = (L_{d,t_i})_{i=1}^n$ of its visited locations $L_{d,t} \in \mathcal{P}$ such that $\forall i: (t_i<t_{i+1}) \land \nexists t':(t_i<t'<t_{i+1}) \land (L_{d,t_i} \neq L_{d,t'}) \land (L_{d,t'} \neq L_{d,t_{i+1}})$. The length of $T_d$ is the number of location transitions $|T_d| := n-1$. A device's \emph{complete trajectory} is the one containing all its locations.
\end{definition}
We define an (abstract) topological map as a directed graph where each location is a vertex and each trajectory of length $1$ is an edge. We call this an \emph{abstract} map as opposed to visualized topological maps.

\begin{definition}[Abstract Topological Map]
An abstract topological map of a WiFi data set $W$ with a partition $\mathcal{P}$ is a directed, edge-weighted graph $G=(V,E)$ with $V=W/L=\mathcal{P}$ and $E \subseteq W/L \times W/L$ such that $(u,v) \in E$ iff there exists a trajectory of length $1$ from location $u$ to $v$. Its edge weights are determined through an edge-weight function $w: E \rightarrow \mathbb{R}^+$.
\end{definition}
Hence, edges are directed and may reflect asymmetric relations between locations.
Additionally to the edges, relations are represented by an edge-weight function, for which we conceptualize several options. Let $\mathcal{T}$ be the multiset of all trajectories of length 1 from all devices in a WiFi dataset $W$. By abuse of notation, we denote $\mathcal{T} = \{T_d \mid |T_d| = 1, d \in D\}$ where the same $T_d$ can occur multiple times in $\mathcal{T}$. We denote the number of occurrences of $T_d$ in $\mathcal{T}$ as $m(T_d)$.
\begin{enumerate}
\item{\emph{Transition Count}: 
$w_{\tau}: (u,v) \mapsto m(u,v)$. 
The number of transitions between the locations by any device. This can be imagined as a "trodden path" which solidifies the more often people walk from $u$ to $v$.}
\item{\emph{Transition Probability}: 
$w_{p}: (u,v) \mapsto m(u,v) / \sum_{w \in V}{m(u,w)}$. 
The number of transitions from $u$ to $v$ relative to the number of transitions from $u$ to any other node. This can be regarded as an estimate of the probability of walking to $v$, given that the current location is $u$ assuming a Markov process.}
\item{\emph{Time Difference}:  $w_{\Delta} : (u,v) \mapsto median(\{t_{i+1} - t_{i} \mid \exists d \in D, j \in \mathbb{N}: t_i = \pi_j(\sigma_d) \land L_{d,t_i} = u \land L_{d,t_{i+1}} = v\})$ where $\pi_j(\sigma_d)$ is the $j$-th timestamp in the motion mode segmentation $\sigma_d$ (cf. \cref{sec:motionmode}). This is the average transition time between locations as a proxy for their distance. The median is used since it is robust against outliers (i.e., extraordinarily fast or slow transitions).}
\item{\emph{Distance in Signal Space}: $w_{\sigma} : (u,v) \mapsto d(u,v)$ for some distance measure $d$, such as the average EMD between the WiFi segments related to $u$ and $v$. Similar to the \emph{time difference}, this estimates distances between real locations.} 
\end{enumerate}
The resulting representation can be visualized through one of many graph-drawing algorithms. In our method, so far, we assumed that the data has been collected in advance and the resulting map reflects all visited locations. However, in a practical scenario, such as a conference, one may wish to calculate and visualize maps in \emph{real-time}. Our method can achieve that by repeating all calculations at different points in time with all data collected up to that moment. 

\section{Experiments}
\label{sec:experiments}
We designed and conducted three experiments at two different venues.
To increase the generalization of our results both venues have
different properties, e.g., one does have multiple floors while the other
does not. In one experiment we controlled the movement while the other is a real-world experiment.
Moreover we employed two different methods to acquire the
ground truth to evaluate our results. The experiment was conducted using signals from the 2.4GHz band without any modification to the pre-installed APs.
We employ different smartphone devices. In particular, we used three HTC One X+, one Samsung Galaxy S8, one Huawei P20, one Honor 8X, three Samsung Galaxy S2 and four Samsung Galaxy S Plus devices. This diversity allowed us to simulate a more realistic scenario.

\subsubsection{SFN Dataset}

During a two day congress held at the ``Schülerforschungszentrum
Nordhessen'' (SFN) we recorded two controlled \emph{walkarounds} by
the first author. For this, all smartphones were attached to the back
of this person, simulating backpack carrying. While walking to and
staying at relevant locations the smartphones collected WiFi data and
sent it to a central server. The locations were presentation booths or
other relevant rooms, altogether twelve distinct places:
\emph{Computer Lab}, \emph{Loft}, \emph{3D-Printing},
\emph{Presentation Room}, \emph{Workshop}, \emph{Working Space},
\emph{Electron Microscope}, \emph{Conservatory}, \emph{Sound Lab},
\emph{Moon Station}, \emph{Laser Lab}, \emph{Chemical Lab}.
The whole building, as depicted in Figure~\ref{fig:sfnmaps}, had four
floors with a varying number of openly accessible rooms.
%
%
Each walkaround had a duration of about one hour. We used 13 devices.
%
For ground truth, we installed ESP32\footnote{\url{https://www.espressif.com/sites/default/files/documentation/esp32-wroom-32_datasheet_en.pdf}} 
modules in all the relevant rooms. These devices were configured as WiFi APs, such that their presence could be detected by the smartphones whenever they were nearby. The ground truth location for a stationary WiFi segment could then be determined through the ESP32 with the highest average signal strength. 

\subsubsection{Office Dataset}
We collected WiFi signals and acceleration data from seven individuals
at our institute (cf.~\cref{fig:officemaps}) during one work week
(i.e., five full days of work time). This data was collected using
eight\footnote{One individual carried two smartphones all the time.}
pairwise different smartphone models. Additionally each participant
carried an RFID badge attached to the chest. Together with stationary
RFID badges we were able to establish a ground truth location. 
%
%
For technical details about this RFID setup we refer the reader to
Schäfermeier et al.~\cite{wifidistances} and Cattuto et
al.~\cite{cattuto2010dynamics}.


\subsection{Measures of Cluster Quality}
We assess the quality of clustering WiFi segments using several
measures that depend on a ground truth. We decided to employ multiple
measures since every measure covers different aspects of accordance to
the ground truth. Any clustering depends on the processing step
\emph{likelihood estimation}~\cref{likeli} and \emph{pairwise distance
  calculation}~\cref{sec:distcalc}, see Figure~\ref{fig:method}. For a
thorough study on this matter we refer the reader to Schäfermeier et
al.~\cite{wifidistances}, from which we use the recommended methods.
Furthermore, in our experiments we fix the clustering method to
HDBSCAN for the reasons we explained in~\cref{sec:clustering}.
The remaining free parameter is the used layout method. Hence, we
calculate the evaluation measures for MDS, t-SNE, and UMAP. Since
these methods have random components, we perform several runs of the
algorithms and report the average scores. We also calculate
the standard deviation to provide an idea of
the achieved stability.

\def \tabang{60}
\begin{table}[b!]
\caption{Clustering evaluation for different layout methods. We present the average scores and standard deviations of each cluster quality measure after ten runs of the experiment. The best scores are printed bold and were determined from the average values with a third decimal place not depicted here.}\label{tab:evaluation}
\centering
\begin{tabular}{p{0.5cm} l l | l l l | l l l}
  \toprule
  &&&\multicolumn{3}{c|}{\bfseries 2D Embedding}
  &\multicolumn{3}{c}{\bfseries 3D Embedding}\\
  \cmidrule{4-9}
  & & \bfseries \rotatebox{\tabang}{Method} &\bfseries\rotatebox{\tabang}{AMI} &
\bfseries \rotatebox{\tabang}{ARI}&\bfseries  \rotatebox{\tabang}{$V$-Meas.} & \bfseries \rotatebox{\tabang}{AMI} & \bfseries \rotatebox{\tabang}{ARI}
  &\bfseries  \rotatebox{\tabang}{$V$-Meas.} \\
\cmidrule{2-9}
\multirow{3}{*}{\bfseries \rotatebox{90}{Day 1}} &
&\textbf{MDS}&.66$\pm$.02&.56$\pm$.02&.75$\pm$.01
                                  &\textbf{.74}$\pm$.01&\textbf{.66}$\pm$.01&\textbf{.82}$\pm$.01\\
&&\textbf{UMAP}&.70$\pm$.01&.53$\pm$.02&.80$\pm$.01&.71$\pm$.01&.54$\pm$.02&.81$\pm$.01\\
&&\textbf{TSNE}&\textbf{.82}$\pm$.00&\textbf{.65}$\pm$.01 &\textbf{.84}$\pm$.01&.50$\pm$.15&.43$\pm$.15&.55$\pm$.15\\
\midrule
\multirow{3}{*}{\bfseries \rotatebox{90}{Day 2}}&
  &\textbf{MDS}&.51$\pm$.03&.52$\pm$.03&.57$\pm$.03
&\textbf{.60}$\pm$.03&\textbf{.56}$\pm$.07&.67$\pm$.02\\
  &&\textbf{UMAP}&\textbf{.60}$\pm$.01&\textbf{.54}$\pm$.04
&\textbf{.72}$\pm$.03&.60$\pm$.01&.53$\pm$.03&\textbf{.71}$\pm$.02\\
  &&\textbf{TSNE}&.59$\pm$.08&.53$\pm$.11&.71$\pm$.08
                                  &.16$\pm$.16&.08$\pm$.11&.21$\pm$.17\\
\bottomrule
\end{tabular}
\end{table}

For a set $X$, let $\mathcal{C}$ and $\mathcal{K}$ be partitions where
the latter is a computed clustering and former is the ground truth.
\begin{inparadesc}
\item[Adjusted Rand index~\cite{adjustedrand}:] Let $a\coloneqq
  |\{\{x,y\}\in{X\choose{2}}\mid \exists C\in \mathcal{C}:\{x,y\}\in
  C\wedge \exists K\in \mathcal{K}:\{x,y\}\in K\}|$, i.e., $a$ is the
  number of element pairs such that both elements are assigned to the
  same cluster within $\mathcal{C}$ as well as $\mathcal{K}$. Dually,
  let $b\coloneqq|\{\{x,y\}\in{X\choose{2}}\mid \nexists C\in
  \mathcal{C}:\{x,y\}\in C\wedge \nexists K\in \mathcal{K}:\{x,y\}\in
  K\}|$. Then $ARI = (RI - \mathbb{E}[RI])/(\max(RI) -
  \mathbb{E}[RI])$ where $RI = (a+b)/{X\choose{2}}$ is the
  \emph{Rand index}.  The term $\mathbb{E}[RI]$ is the expected
  value of $RI$.  The Rand index is adjusted such that a random
  labeling will get an $ARI$ close to zero and the best possible
  score is 1.0.
\item[Adjusted mutual information~\cite{adjustedmutualinfo}:] Let
  $P_{\mathcal{C}} (C)$ denote the probability of a random sample
  $x\in X$ falling into cluster $C$ in the partition
  $\mathcal{C}$. Respectively, $P_{\mathcal{C}\mathcal{K}}(C,K)$
  denotes the probability of a sample belonging to cluster $C$ in
  $\mathcal{C}$ and to $K$ in $\mathcal{K}$. Based on that the
  \emph{mutual information} of $\mathcal{C}$ and $\mathcal{K}$ is
  defined via $MI(\mathcal{C},\mathcal{K})\coloneqq
  \sum_{C\in\mathcal{C}}
  \sum_{K\in\mathcal{K}}P_{\mathcal{C}\mathcal{K}}(C,K) \log
  (P_{\mathcal{C}\mathcal{K}}(C,K)/(P_{\mathcal{C}}(C)
  P_{\mathcal{K}}(K)))$. In the following, let $H(\mathcal{C})$ denote
  the Shannon entropy of partition $\mathcal{C}$. Then the number $AMI
  = (MI(\mathcal{C},\mathcal{K}) -
  \mathbb{E}[MI(\mathcal{C},\mathcal{K})])/(\max(H(\mathcal{C}),
  H(\mathcal{K})) - \mathbb{E}[MI(\mathcal{C},\mathcal{K})])$ is the
adjusted mutual information, which uses the expected value
$\mathbb{E}[MI(\mathcal{C},\mathcal{K})]$ given partitions of the same
size as $\mathcal{C}$ and $\mathcal{K}$.
\item[Cluster Homogeneity~\cite{vMeasure}:] This measure reflects to
  what extent elements that are assigned to a cluster belong to the
  same single class, with respect to the ground truth. The
  \emph{cluster homogeneity} is defined by
  $h(\mathcal{C},\mathcal{K})\coloneqq 1 -
  H(\mathcal{C}|\mathcal{K})/H(\mathcal{C})$, where
  $H(\mathcal{C}|\mathcal{K})$ denotes the conditional entropy of the
  class assignment $\mathcal{C}$ given the clustering $\mathcal{K}$.
\item[Cluster Completeness~\cite{vMeasure}:] This measure reflects to
  what extent elements from the same ground truth cluster in
  $\mathcal{C}$ are assigned to the same cluster in
  $\mathcal{K}$. Based on that idea is the \emph{cluster completeness}
  $c(\mathcal{C},\mathcal{K})\coloneqq 1 -
  (H(\mathcal{K}|\mathcal{C}))/(H(\mathcal{K}))$.
\item[$V$-measure~\cite{vMeasure}:]
Based on the harmonic mean of cluster homogeneity and cluster
completeness the \emph{$V$-measure} is defined by
$V(\mathcal{C},\mathcal{K}) = 2\cdot h(\mathcal{C},\mathcal{K})\cdot c(\mathcal{C},\mathcal{K})/(h(\mathcal{C},\mathcal{K}) + c(\mathcal{C},\mathcal{K}))$.
Analogously to $F$-measure for classification, the values of both measures must be high in order to achieve a high value for the $V$-measure. 
\end{inparadesc}

\subsection{Experimental Evaluation of Cluster Quality}

\begin{figure}[b!]
    \includegraphics[trim=0 0 30 0, clip,width=0.5\columnwidth]{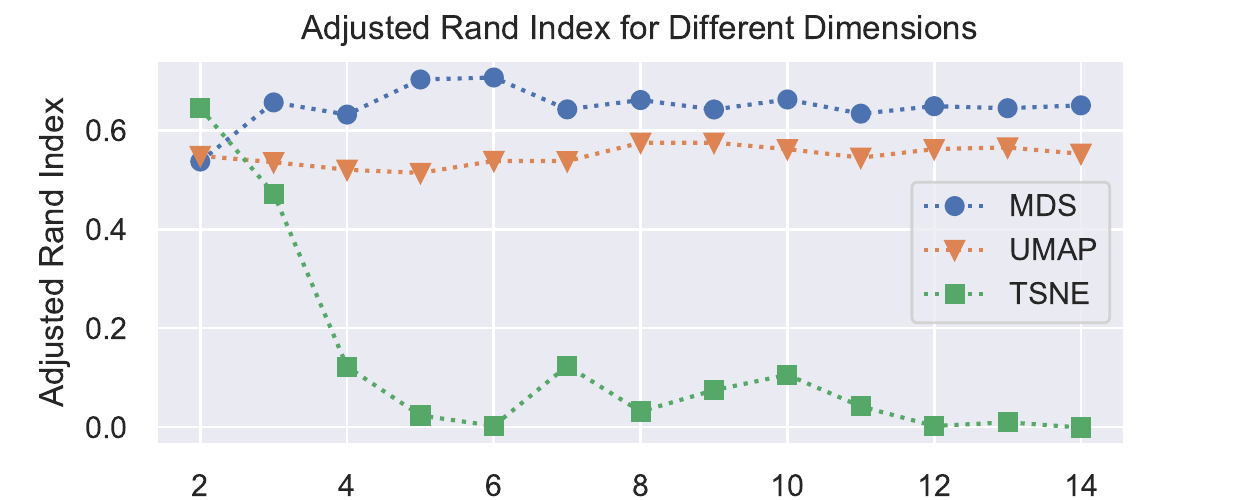}
  \includegraphics[trim=0 0 30 0, clip,width=0.5\columnwidth]{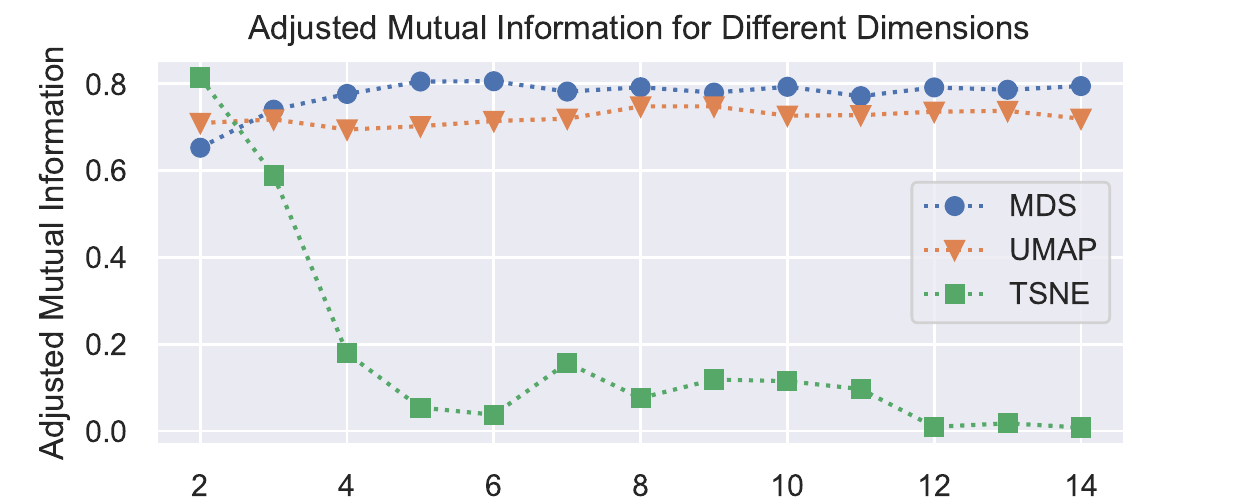}\\ \vspace{-4.5ex}
\begin{center}\includegraphics[trim=0 0 30 0, clip,width=0.5\columnwidth]{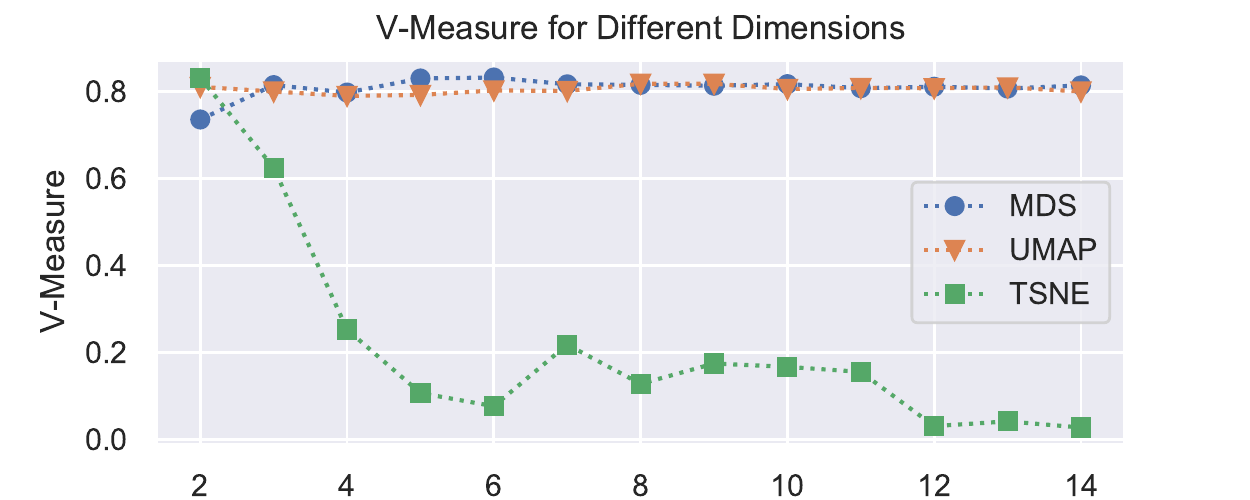} \end{center}
  \caption{Performance of layout methods for dimensions two to fourteen.}\label{fig:score_comparison}
\end{figure}

\paragraph{Comparison of 2d embeddings}
In our first experiment, we applied all layout algorithms (MDS, t-SNE,
UMAP), with a dimension of two, and the subsequent clustering
algorithm (HDBSCAN) to the SFN data set. Since our overall goal is to
create topological maps, the initial choice for dimension two is	
natural. We chose SFN, since the ground truth labels for this data set
are particularly trustworthy, as they were determined through an
additional manual log. To obtain evidence for the stability of the
results we conducted each run ten times. We report
arithmetic means and standard deviations for the cluster-quality measures in~\cref{tab:evaluation}.

We observe that both UMAP and TSNE give considerably better results
than MDS for all of the evaluation measures. 
Furthermore we note that TSNE performs better than or equal to
UMAP in all experiments.
Comparing the scores of day 1 with day 2, notably they are higher on the first day. 
We may note that, even though these values are not comparable due to the different walks performed,
the same consequences regarding the compared layout methods apply.

In a visual examination of the plots we found: UMAP leads to plots
where points in a cluster are concentrated close together, almost
falling into a single point. For TSNE we still get clusters
that are clearly distinguishable. However, the points within each
cluster are not as densely concentrated.  We may note
that a more dense plot of a cluster does not necessarily imply a
better clustering result with respect to the ground truth, as can be inferred from~\cref{tab:evaluation}.

\paragraph{Comparison of 3d and higher-dimensional embeddings}
In the second experiment on the SFN data set, we studied higher-dimensional embeddings.
We specifically report our findings on day 1
(see \cref{fig:score_comparison}), for which we conducted one run per
dimension and method. We omit the report on day 2, since results were
similar.

First, we find that UMAP and MDS already score close to the best value
for a low number of dimensions, i.e., three or four. A further
increase in dimension causes only a slight increase of the
scores. Hence, it seems that the three dimensional embedding space is
sufficient to capture the WiFi signal space that arises from the
underlying three-dimensional space. Second, although TSNE achieved the
best scores for dimension two (\cref{tab:evaluation}) they decrease
for higher dimensions. 
In this case, MDS is the method that gives the best results. 
%

These results motivated us to settle for dimension three for the rest
of this investigation, which we also report in~\cref{tab:evaluation},
We find that MDS in 3d performs similar well as TSNE in 2d and better than 2d MDS.
TSNE, as already seen in \cref{fig:score_comparison} performs worse in 3d.
MDS performs better than UMAP in most cases, although scores are often similar.
Overall, from our experiments, we concluded that multidimensional scaling is the method of choice for the layout step.
We also concluded that using the dimension number of the underlying physical space should be used, i.e., for a venue with several floors, we use 3d embeddings.

\subsection{Generated Maps}
\begin{figure}[t!]
\begin{center}
\includegraphics[trim=65 20 45  55, clip, width=0.49\columnwidth]{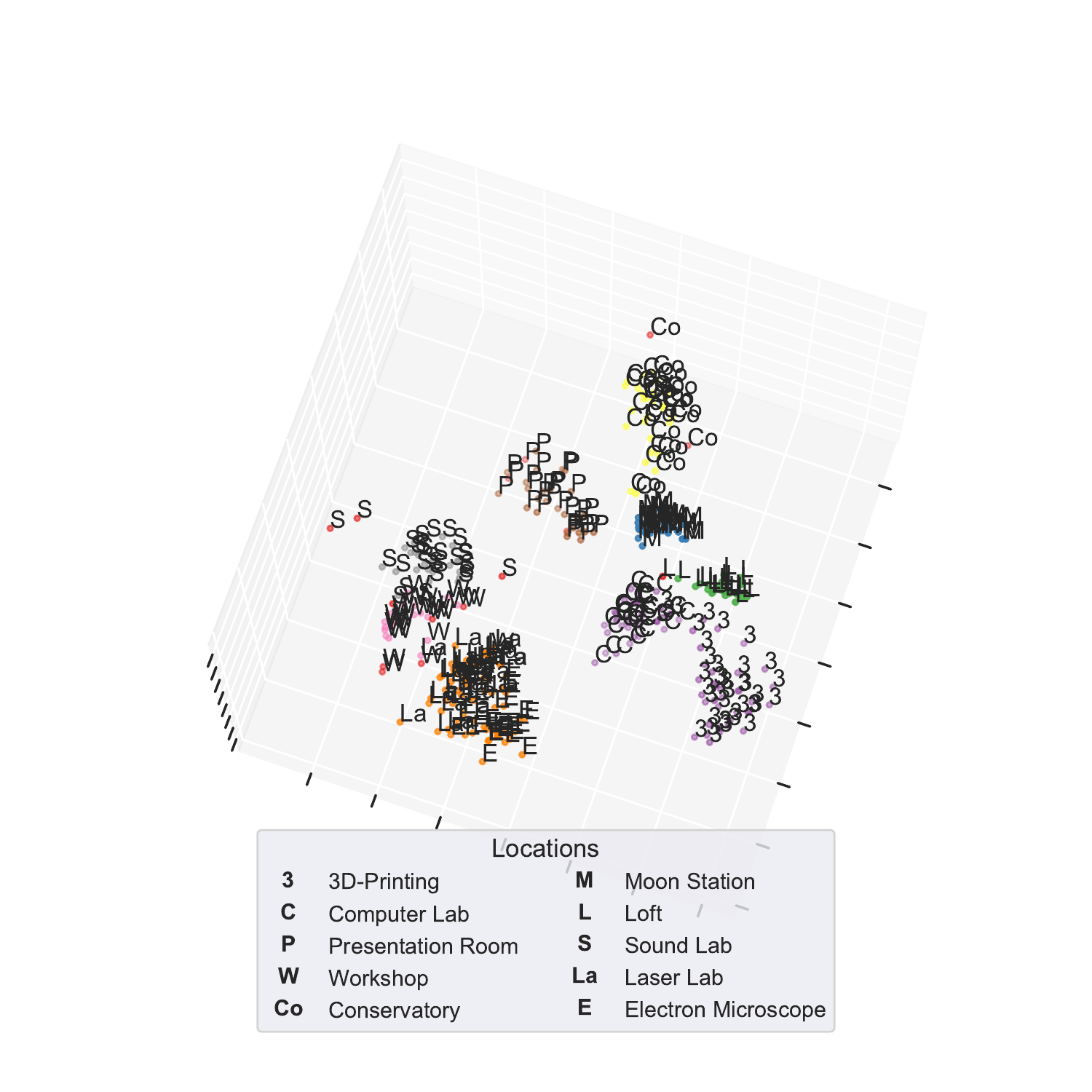}
\includegraphics[trim=65 20 45  55, clip, width=0.5\columnwidth]{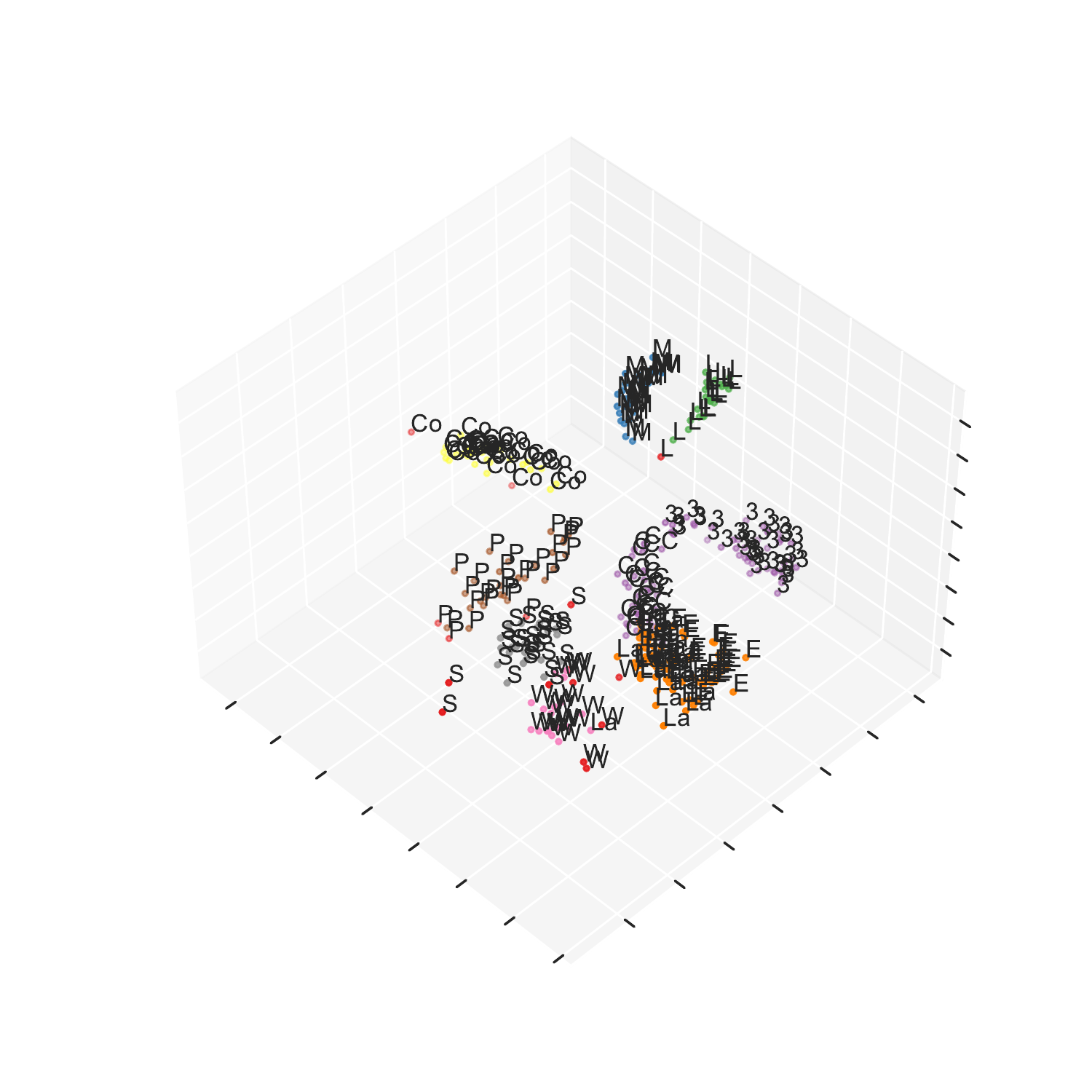}\\
\vspace{0.25cm}

\def \floorplanscaling {0.5}
\scalebox{.49}{
\includegraphics[width=\floorplanscaling\columnwidth]{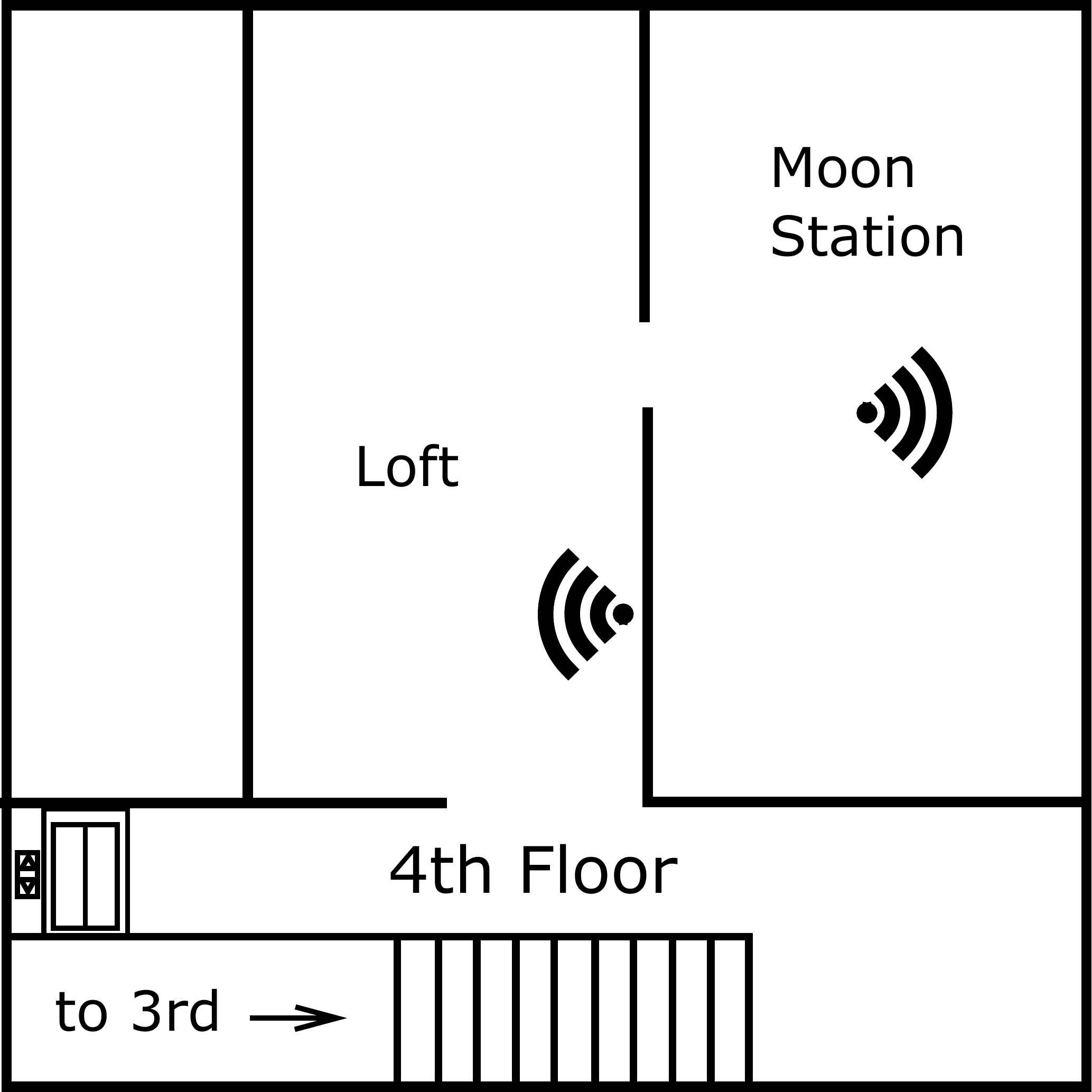}
\includegraphics[width=\floorplanscaling\columnwidth]{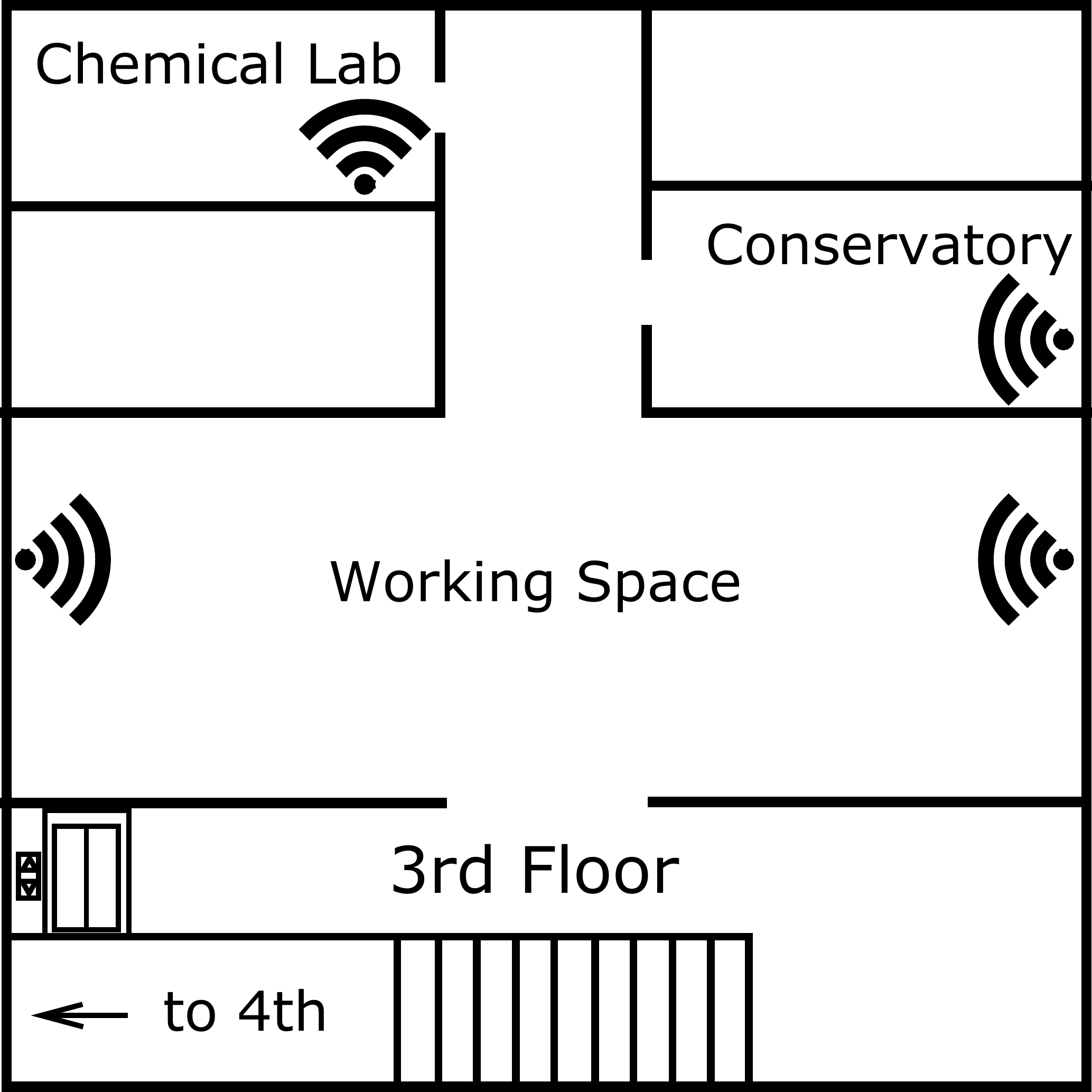}
\includegraphics[width=\floorplanscaling\columnwidth]{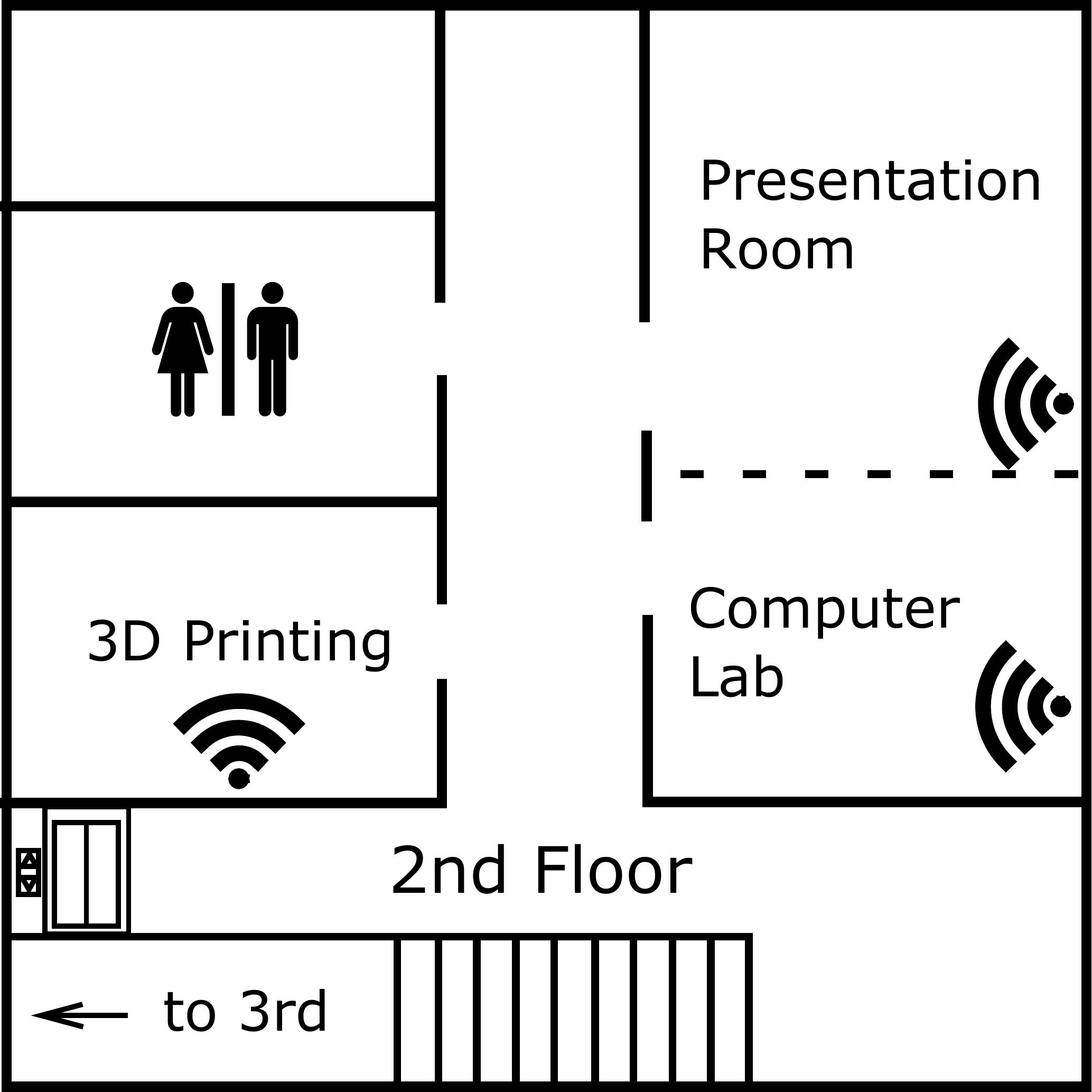}
\includegraphics[width=\floorplanscaling\columnwidth]{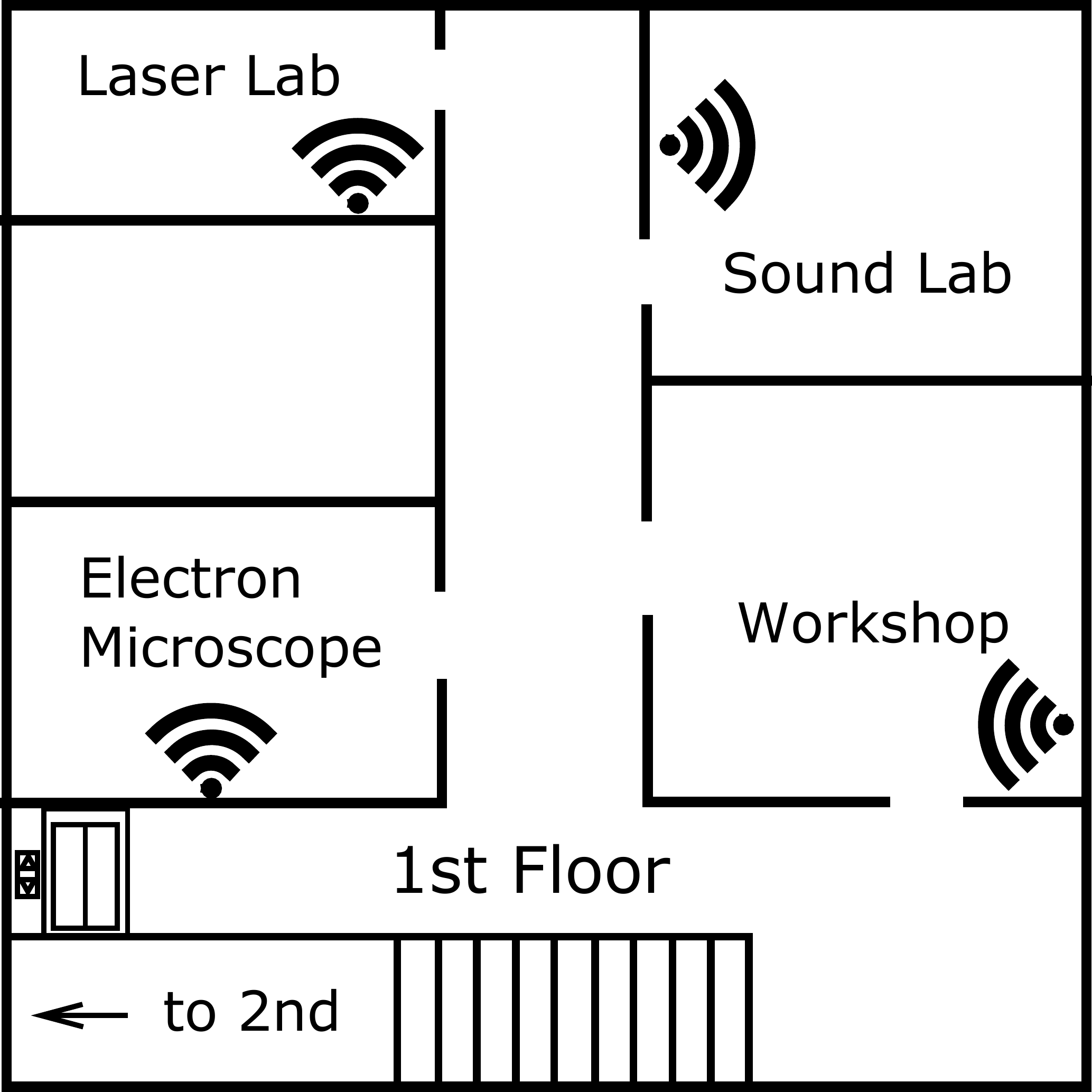}
}
\end{center}
\caption{\textbf{Top:} SFN wifi segment layout and clustering for day 2. Each segment is annotated with its ground truth location and colored by its cluster. Outliers are red. Two locations were omitted in the experiment since access was currently not possible. Best viewed in color;
\textbf{Bottom:} Floor plan of the venue where the SFN data set was recorded. In this figure, locations of self-installed WiFi access points (ESP32 modules) are given.}\label{fig:sfnmaps}
\end{figure}

Based on our method we generated topological maps for the SFN and office data set using MDS as depicted in \cref{fig:officemaps,fig:sfnmaps,fig:sfnmaps}. For both cases, we show floor plans of the building and the embedded and clustered WiFi segments. For the office data set, we use 2d embeddings since the experiment was conducted in a single-floor building. As can be seen, the ground truth locations can be recovered. Moreover, we find that the topological relations between locations are accurately captured. In detail, one can infer the physical floor plan from this plot to some extent. However, we find that in a few cases neighboring locations fall together into a single cluster. This can be attributed to the underlying WiFi signals being similar when being close to the wall between two rooms. 
We also observed that the layout of the WiFi APs within buildings contributes to the performance.
 
\begin{figure}[t!]
\begin{center}
\includegraphics[width=0.42\columnwidth]{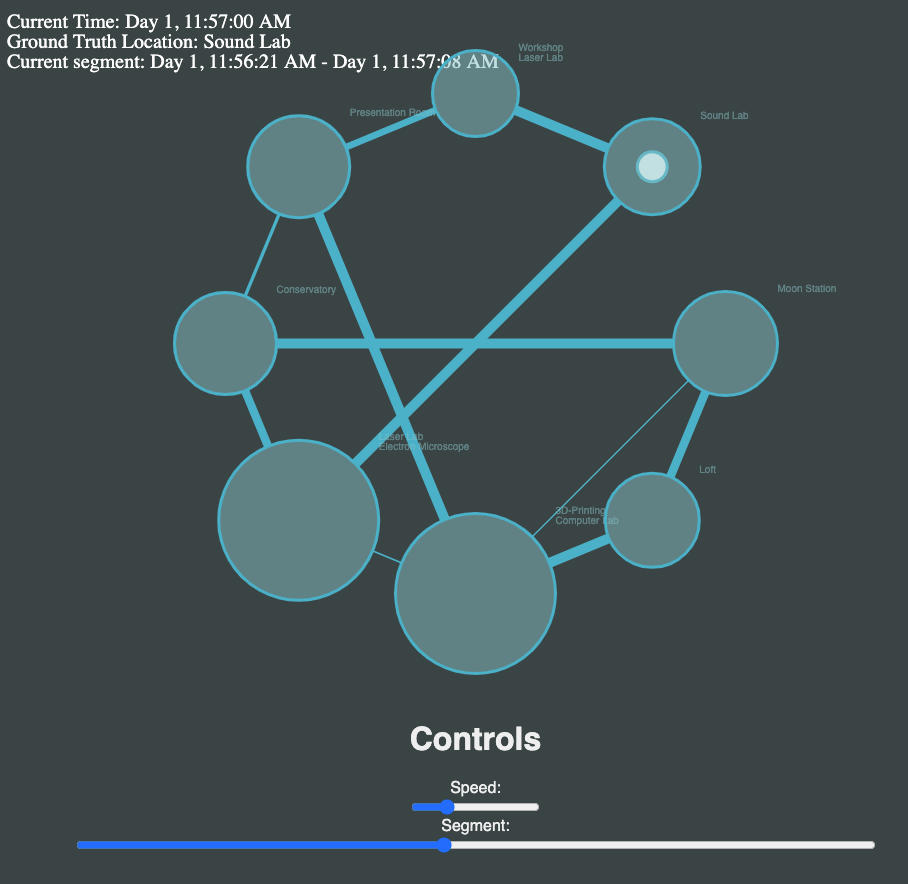}
\end{center}
\caption{Topological map demo generated for the SFN data set.}\label{fig:topomap}
\end{figure}

We created a demo application for the depiction of topological maps during events. The application displays the recognized locations and constructs a topological map of them~(\cref{fig:topomap}). Furthermore, it allows to display and track the locations of devices over time and shows the transition probabilities.
Locations, which in our model are represented as graph nodes, are depicted as circles. The thickness of a line between the nodes represents the edge weight (cf. \cref{sec:graphrepresentation}). The locations of devices in this map are visualized as dots inside the nodes, in our example there is only one device. 



\subsection{Location Log Comparison of Segments and Ground Truth}
Our mapping procedure allows us to extract location logs for devices as depicted in \cref{tab:locationlog}, in which each row represents a stationary WiFi segment over time. The \emph{location} column here contains the ground truth locations (determined through the ESP32 controllers). We used these location logs as a compact representation of the results of our method that can, e.g., be compared to manual logs.

As an example, we gathered a manual location log at the office venue with forty-five log entries, from which forty-two entries were inferred correctly through our method. In one case, a single entry in our log was recognized as two segments. In two cases, the determined ground truth annotation was mistaken for an adjacent room, since the walls here were very thin and the ESP32 controller in the same room was slightly rotated. Altogether, although we do not systematically quantify these results, we found our motion mode segmentation as well as our system for determining the ground truth to function reasonably well.

\begin{table}[t!]
\caption{Illustrative example for a location log of a device.} \label{tab:locationlog}
\begin{center}
\begin{tabular}{p{2cm} p{3cm} p{3cm} p{2cm}}
\toprule
\bfseries Location & \bfseries Start & \bfseries End & \bfseries Duration\\
\midrule
Loc. A & Day 1 09:26:04 & Day 1 09:26:44 & 00:00:39 \\
Loc. B & Day 1 09:26:50 & Day 1 09:27:01 & 00:00:11 \\
Loc. C & Day 1 09:27:06 & Day 1 09:27:12 & 00:00:06 \\
Loc. B & Day 1 09:27:17 & Day 1 09:27:35 & 00:00:18 \\
Loc. D & Day 1 09:27:47 & Day 1 09:28:01 & 00:00:14 \\
\bottomrule
\end{tabular}
\end{center}
\end{table}

\subsection{Signal Strength Distributions at Locations}
We combine all segments from a cluster into an overall RSSI likelihood by averaging over the estimated densities for each WiFi segment (\cref{fig:locdistributions}).
An alternative method would have been a kernel density estimation over the samples from the WiFi segments. We favor the first method to avoid single, very long segments dominating the overall estimation. Our choice leads to slightly different positions being incorporated in the overall estimate. 
Based on these representations of our final locations, we can achieve localization for new WiFi observations in our abstract topological map. More specifically, for a WiFi observation, i.e., a vector of RSSI values over all APs, we can achieve localization through maximum likelihood~\cite{horus2005}.
The inferred locations of devices can be, e.g., used for creating a live view of a topological map at some venue, such as a conference or workshop.

\subsection{Movement Analysis}
In our experiments we found that people in an office scenario are stationary for more than $95\%$ of the time. We figure that for the envisioned applications, such as conferences and workshops, this observation will be replicated to some extent.
Hence we base our analyses on stationary segments. We claim that the information conveyed accurately represents the topological locations and relations. This claim is supported by the fact that movement is mostly done between locations of interest while locations visited during movement are mostly hallways.



\section{Outlook}

We presented a simple and lightweight method to derive meaningful topological maps
from WiFi data. In particular, we demonstrated that this is possible without any modifications to the infrastructure of a venue.
We showed in experiments conducted at two different venues that we can identify physical locations and recover their topological relations. 
Our study shows, that overall multidimensional scaling performs best for the layout and clustering task. We found evidence that using the dimension of the venue, i.e., 3d for multistory buildings and 2d for single floors, leads to favorable results. Hence, we conjecture that the intrinsic dimension of the data reflects the physical dimension to some extent. 
Furthermore, in our experiments, it transpired that the employed methods for motion mode segmentation, likelihood estimation and distance computation are suitable for the localization and mapping problem. Our mapping process relies on the clustering method HDBSCAN, which turned out to be robust, as reflected by low standard deviations of the cluster quality measures. Moreover, apart from few exceptions, it reliably determined the correct number of physical locations. 
We showed a suitable graph representation for our topological maps. As a notable feature, we presented four different weight functions that reflect different aspects of distances or user transitions between locations. We envision that these weights will be useful for the analysis of social patterns in the realm of events, such as conferences or fairs. Also these edge weights can convey interesting information, when visualized within the final result of our method, i.e., topological maps.

We see many connections of our analysis methods to current research in the area of ubiquitous computing. The next logical step is to enhance our maps with location semantics, e.g., recognizing a registration desk at a conference venue. One can also imagine to automatically derive topical descriptions~\cite{schafermeier2020topic} for locations based on volunteered publication data from the conference participants.
As a limitation, we noticed that the device diversity problem could be addressed through more sophisticated methods. The graph drawings could be improved further by adopting, e.g., temporal graph drawing algorithms, which preserve the mental map of users~\cite{graphdrawingBrandes,
  graphdrawingTemporal, graphdrawingDynadag}. Finally, a particularly interesting application would be to analyze user trajectories, with methods such as HypTrails~\cite{hyptrails}.

\printbibliography

\appendix

\section{Complementary Visualizations}
\begin{figure}[H]
  \begin{center}
    \includegraphics[trim=30 5 50 20, clip,width=0.75\columnwidth]{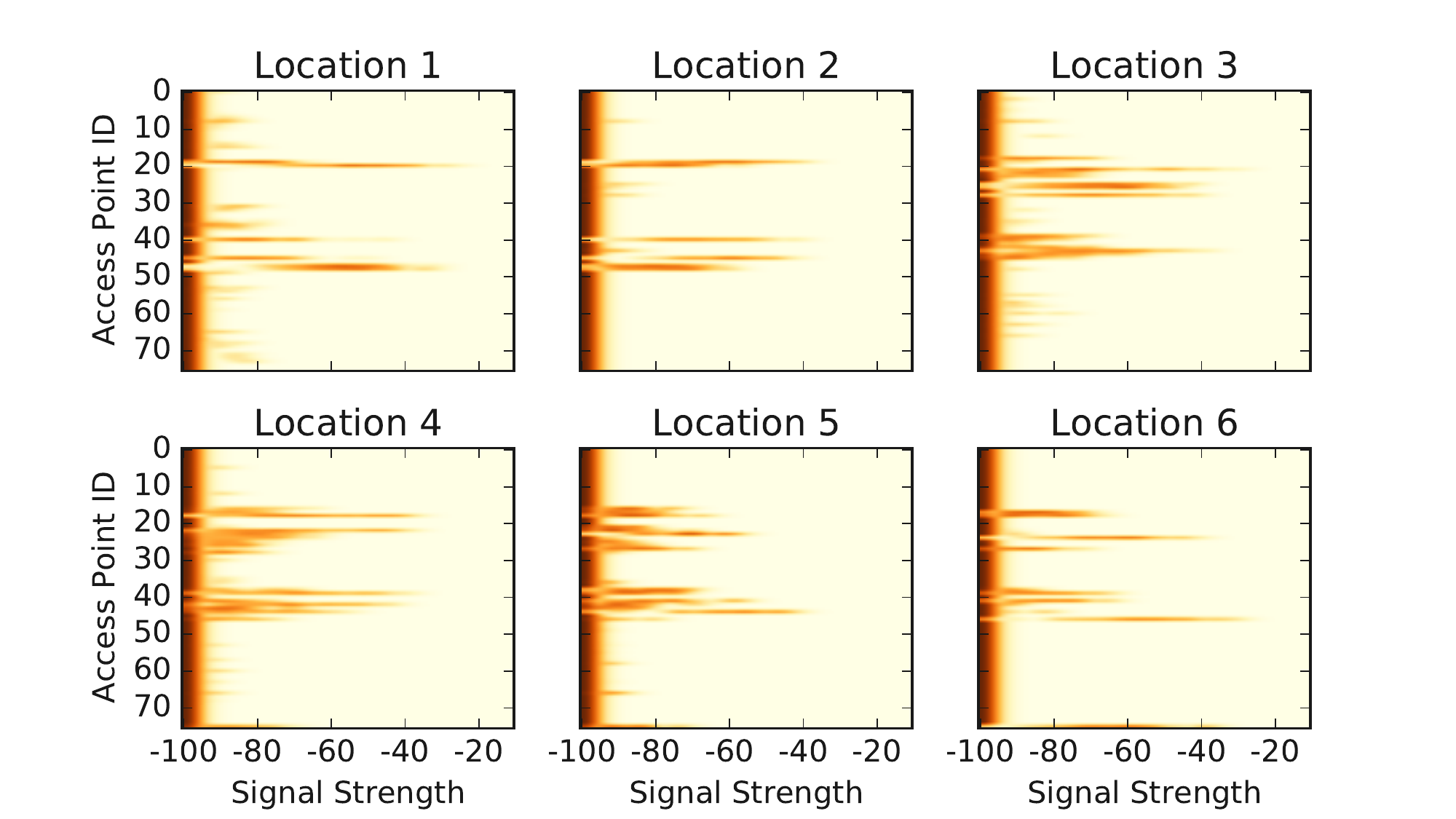}    
  \end{center}

  \caption{Signal Strength Distributions at Locations. Each subplot depicts the RSSI likelihood given a discovered location. Each row on the y-axis of a subplot represents   the signal strength distribution an access point. Colors indicate estimated probabilities. Plots are given for an exemplary subset of the locations discovered by our method in the SFN walkaround 1 data set.}\label{fig:locdistributions}
\vspace{-1.1cm}
\end{figure}

\begin{figure}[H]
  \centering
  \includegraphics[angle=0,width=0.20\columnwidth, trim=140 0 140 40, clip]{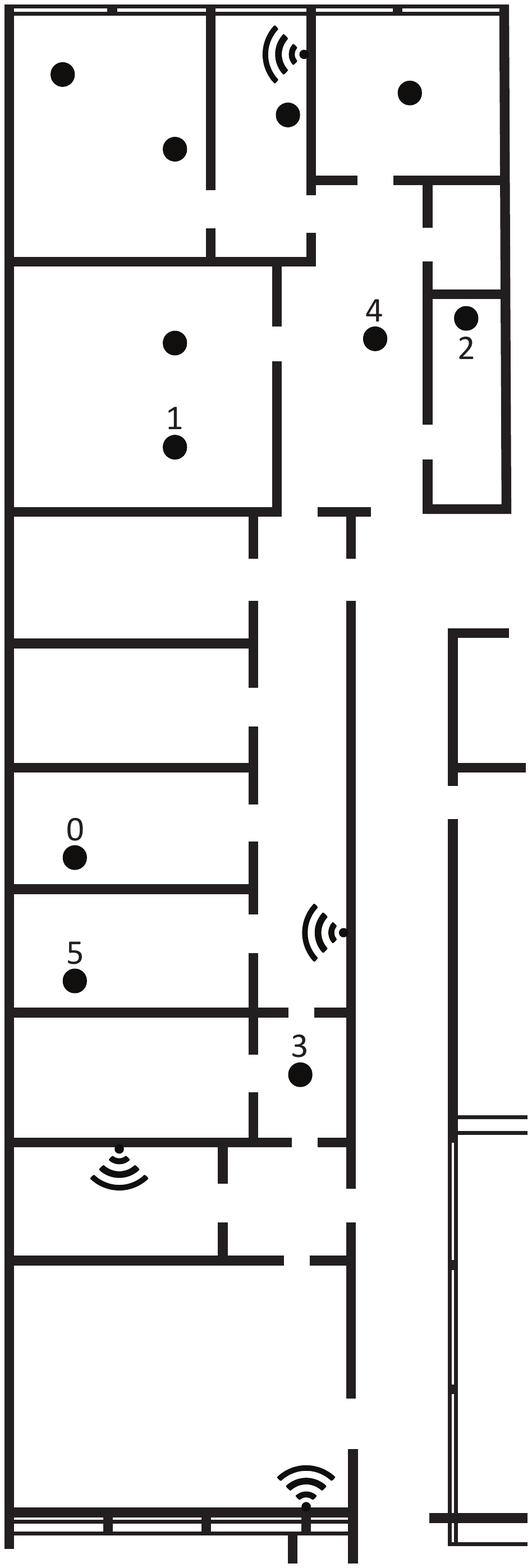}  
  \includegraphics[trim=120 160 130  180, clip,width=0.63\columnwidth]{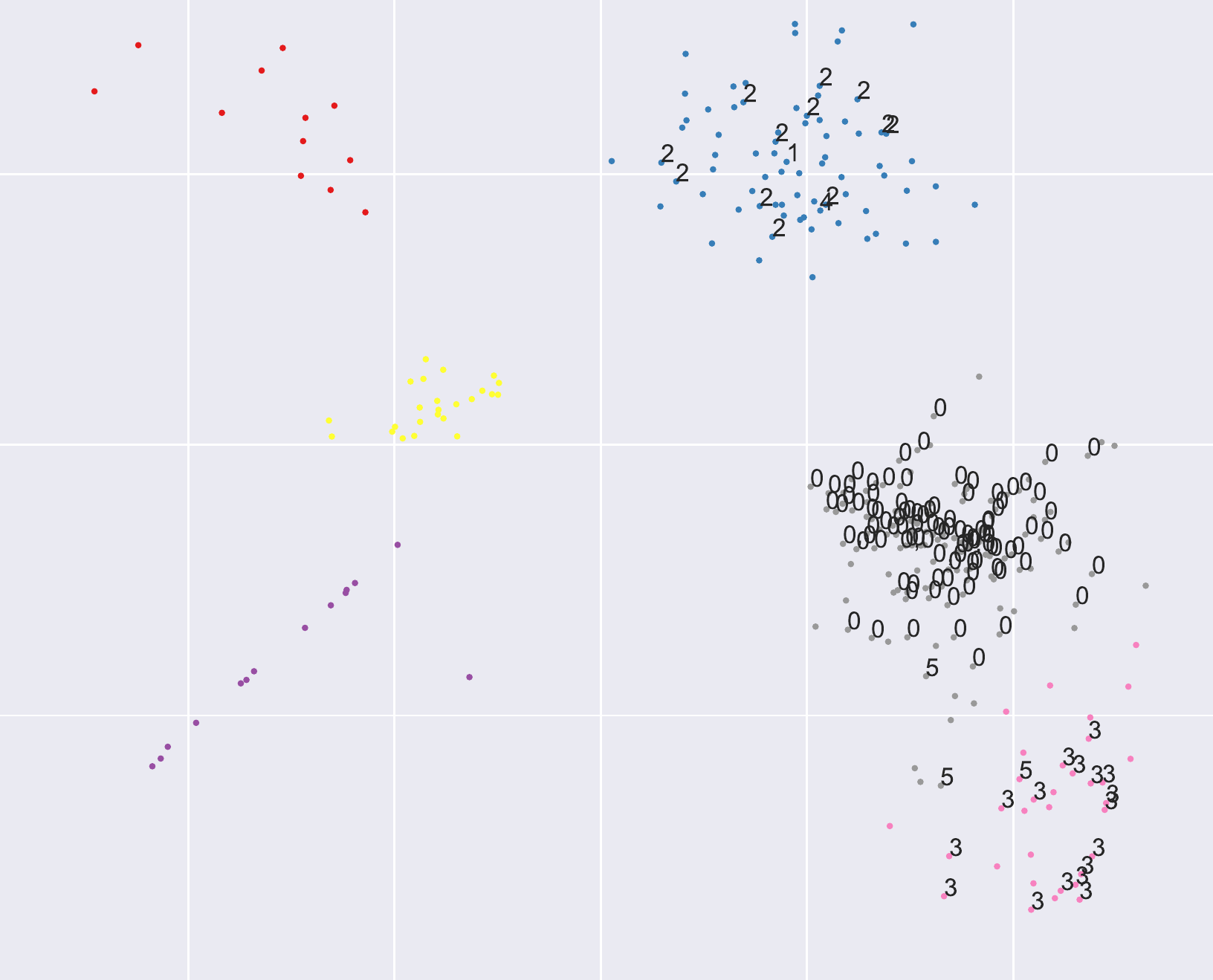}
  \caption{\emph{Left:} Office floor plan, incl. AP locations and
    stationary RFIDs (dots). Relevant RFIDs are annotated by
    numbers. \emph{Right:} Calculated MDS coordinates per
    device. Points represent stationary WiFi segments and are
    clustered through HDBSCAN (one color per cluster). If contacts
    with stationary RFIDs were registered we annotated points with the
    ground truth (see floor plan). Unlabeled clusters:
    canteen, neighboring grocery store, lavatory. Best viewed in
    color.}\label{fig:officemaps}
\end{figure}

\end{document}